\documentclass[pra, 10pt, twocolumn, letter, floatfix]{revtex4}

\usepackage{amsmath}
\usepackage{amsfonts}
\usepackage{amssymb}
\usepackage{graphicx}
\usepackage{rotating}
\allowdisplaybreaks
\usepackage{setspace}
\usepackage{bm}

\begin{document}

\title{A Jastrow factor coupled cluster theory for weak and strong electron correlation}

\author{Eric Neuscamman\footnote[1]{Electronic mail: eric.neuscamman@gmail.com}}
\affiliation{Department of Chemistry, University of California, Berkeley, California 94720, USA}

%

\begin{abstract}
We present a Jastrow-factor-inspired variant of coupled cluster theory
that accurately describes both weak and strong electron correlation.
Compatibility with quantum Monte Carlo allows for
variational energy evaluations and an antisymmetric
geminal power reference,
two features not present in traditional coupled cluster that facilitate
a nearly exact description of the strong electron correlations
in minimal-basis N$_2$ bond breaking.
In double-$\zeta$ treatments of the HF and H$_2$O bond dissociations,
where both weak and strong correlations are important,
this polynomial cost method proves more accurate than either traditional
coupled cluster or complete active space perturbation theory.
These preliminary successes suggest a deep connection between the
ways in which cluster operators and Jastrow factors encode correlation.
\end{abstract}

\maketitle




Predicting the effects of electron correlation remains one of the
most challenging aspects of theoretical chemistry.
Weak correlations, which include dispersion interactions and 
the effects of electron coalescence cusps, are small in magnitude
and can be described by mean-field-based methods like
coupled cluster (CC) theory. \cite{BARTLETT:2007:cc_review}
The strong correlations in molecules with small gaps
require methods based on other assumptions,
such as active space configuration interaction \cite{Helgaker_book}
or the density matrix renormalization group. \cite{White:1992:dmrg,Chan:2008:dmrg_chem_review,Legeza:2003:dmrg_lif}
However, making accurate predictions in many important chemical systems
(e.g.\ the oxygen evolving complex \cite{Yanai:2013:oec})
requires describing both weak and strong correlations simultaneously.
This difficult goal has attracted much recent research, including
developments in multi-reference CC (MRCC),
\cite{Bartlett:2012:cr_mrcc,Chan:2010:mr_dmrg,Neuscamman:2010:ct_review,Kohn:ic_mrcc_pilot:2011,Kohn:ic_mrcc_pt:2012}
distinguishable CC, \cite{Manby:dist_cc:2013}
quasi-variational CC, \cite{Knowles:2012:quasi_var_cc,Knowles:2012:qvcc_benchmark}
configuration interaction density functional theory, \cite{VanVoorhis:2007:ci_dft,VanVoorhis:2009:ci_dft_barriers}
projected quasiparticle theory, \cite{Scuseria:2011:projected_hfb,Scuseria:2012:projected_hf}
and full configuration interaction quantum Monte Carlo (FCI-QMC). \cite{Alavi:2009:fciqmc,Alavi:2010:fciqmc_initiator}
In the present study, we merge the strengths of CC theory, QMC Jastrow factors (JFs),
and pairing functions to produce a remarkably compact ansatz that accurately
describes both weak and strong correlations.

CC theory encodes correlation using energetic structure:
a reference configuration is mixed with excited
configurations in a size extensive way, with higher excitations
becoming successively less important due to their higher energies.
This energetic hierarchy lets CC truncate its excitation rank
to doubles or triples while maintaining an excellent description of
weak correlation.
CC's poor performance in strongly correlated systems
(doubly excited states, transition metal compounds, dissociating bonds, etc.),
on the other hand,
is primarily a concequence of its mean-field reference, although recent work \cite{Knowles:2010:vcc}
shows its non-variational nature also plays a significant role.

JFs encode correlation using spatial structure via symmetric scalar functions
of electron positions. \cite{FouMitNeeRaj-RMP-01}
While this description is typically less sophisticated than CC, compatibility with QMC
allows diffusion Monte Carlo (DMC)
\cite{FouMitNeeRaj-RMP-01}
to fill in the details.
Indeed, real space Jastrow-Slater DMC provides an accuracy
similar to CC when modeling weak correlations. \cite{Alfe:2012:water_dft_dmc,Mitas:2013:non_cov_qmc}

JFs can also be constructed in Hilbert space (HS), encoding correlation via functions of the
orbital occupation numbers in a localized one-electron basis.
Examples include Gutzwiller factors, \cite{Gutzwiller:1963:gf}
their recent generalization to local particle number projections, \cite{Neuscamman:2012:sc_jagp}
and some tensor networks. \cite{CHANGLANI:2009:cps,NEUSCAMMAN:2011:pcps,MEZZACAPO:2009:entangled_plaquettes,VERSTRAETE:2010:cgtn}
Note that while real space JFs perform a similar role as explicit correlation techniques in satisfying
cusp conditions, the HS-JFs used here do not, as they live entirely within the gaussian orbital basis defining the HS.
In either real space or HS, JFs
suppress configurations in which electrons are too close to each other,
encoding correlation in a more direct (although not necessarily
superior) manner than CC's configuration interaction of
energetically stratified excitations.

This report introduces a cluster-Jastrow (CJ) operator that is
simultaneously a HS-JF and a simplified CC doubles operator.
Its compatibility with QMC allows for variational energy evaluations
at polynomial cost and for a strongly correlated antisymmetric geminal power (AGP)
\cite{Scuseria:2002:hfb,Sorella:2003:agp_sr,Sorella:2004:agp_sr,Scuseria:2011:projected_hfb,Neuscamman:2012:sc_jagp,Neuscamman:2013:jagp}
reference.
Preliminary results in small-molecule bond breaking show that the
CJ operator is a good approximation to CC doubles and that the combined
CJAGP ansatz delivers accurate descriptions of both weak and strong
electron correlation, at least in the relatively small basis sets
accessible to our pilot implementation.
These findings suggest that the traditional views of how JFs and
cluster operators encode correlation may be two sides of the same coin
and motivate further research into their efficient combination.


The CJAGP can be written as a CC ansatz,
\begin{align}
|\Psi\rangle = \exp(\hat{T}) |\Phi\rangle,
\label{eqn:cc_ansatz}
\end{align}
where $\hat{T}$ is an excitation operator and $|\Phi\rangle$ is the reference.
Although the CJ could be generalized to higher excitations,
doubles are typically the most important in chemistry and for now
we restrict ourselves to these,
\begin{align}
\label{eqn:doubles}
\hat{T} &= \sum_{\sigma,\tau \in \{\uparrow,\downarrow\}} \hspace{1mm} \sum_{i,j,k,l\in[1,N]} \hspace{1mm} T_{\hspace{0.8mm}i_\sigma j_\tau}^{k_\sigma l_\tau}
                   a^\dag_{k_\sigma} a_{i_\sigma} a^\dag_{l_\tau} a_{j_\tau},
\end{align}
where $N$ is the number of spatial orbitals and $a^\dag_{k_\tau}$ creates a spin-$\tau$ electron in orbital $k$.
By allowing all indices to range over all orbitals, we expand the typical CC excitation manifold 
in a way reminiscent of generalized CC \cite{Nooijen:2000:gcc}, which is appropriate as
the AGP reference lacks any concept of occupied or virtual orbitals.

The somewhat cumbersome doubles operator is often simplified using locality-based tensor decompositions
\cite{Yang:2012:osv_cc,Pulay:1985:pao,Taube:2008:fno,Krylov:2010:fno_eom,Neese:2011:pno_open_shell}
of the four-index amplitudes $T_{\hspace{0.8mm}i_\sigma j_\tau}^{k_\sigma l_\tau}$.
Here we introduce a new decomposition that makes $\hat{T}$ equivalent to a QMC-compatible HS-JF,
enabling variational energy evaluations and an AGP reference.
We define
\begin{align}
\label{eqn:doubles_decomp}
T_{\hspace{0.8mm}i_\sigma j_\tau}^{k_\sigma l_\tau} &= \sum_{pq} U^{*}_{ip} U_{kp} J_{p_{\sigma}q_{\tau}} U^{*}_{jq} U_{lq}
\end{align}
where $\bm{J}$ is a $2N\times 2N$  matrix of JF weights and
\begin{align}
\label{eqn:antiherm_form}
\bm{U} &= \exp\left(\bm{K}-\bm{K}^\dag\right)
\end{align}
is a unitary one-particle rotation of HS,
\begin{align}
\label{eqn:basis_rot}
\hat{U}^\dag a_{i_\sigma} \hat{U} = \sum_{r} U_{ir} a_{r_\sigma},
\end{align}
under which $\hat{T}$ becomes a HS-JF,
\begin{align}
\label{eqn:doubles_as_jast}
\hat{J} \equiv \hat{U}^\dag \hat{T} \hat{U} & = \sum_{\sigma,\tau \in \{\uparrow,\downarrow\}} \hspace{1mm} \sum_{pq} \hspace{1mm} J_{p_{\sigma}q_{\tau}}
                       a^\dag_{p_\sigma} a_{p_\sigma} a^\dag_{q_\tau} a_{q_\tau}.
\end{align}
Note that our decomposition's similarity to tensor hypercontraction \cite{Martinez:2012:thc_correlated}
helps explain its efficacy, although here $\bm{U}$ is $N\times N$, unitary, and optimized variationally.

In addition to the cluster operator, we specify the AGP reference using its pairing matrix $\bm{F}$, the vacuum $|0\rangle$, and the number of electrons $M$,
\begin{align}
|\Phi\rangle
= \left( \sum_{pq} F_{pq} a^\dag_{p\uparrow} a^\dag_{q\downarrow} \right)^{M/2}|0\rangle.
\label{eqn:agp}
\end{align}
Note that as the AGP is not related to any particular determinant, the present method differs from some MRCC approaches in that there is no dependence on
a reference determinant.
In total, we see that CJAGP contains only $6N^2$ variables, namely the elements of $\bm{J}$, $\bm{K}$, and $\bm{F}$.

The HS Jastrow AGP (JAGP) \cite{Neuscamman:2013:jagp} is a special case of CJAGP obtained by
choosing the local orbitals in Eq.\ (\ref{eqn:basis_rot}) as the $S^{-1/2}$ Lowdin localization of the atomic orbitals.
JAGP provides a qualitatively correct treatment of strong correlation, \cite{Neuscamman:2013:jagp}
and here we hope that the CC character of the orbital-optimized HS-JF will
allow CJAGP to treat weak correlation as well.


We optimize CJAGP variationally with the function
\begin{align}
\label{eqn:cjagp_energy}
E(\bm{K}) = & \hspace{1mm} \substack{\mathrm{min} \\ \bm{J} \bm{F}} \hspace{1mm}
              \big[ \langle\Phi| e^{\hat{T}^\dag} \hat{H} e^{\hat{T}} |\Phi\rangle / \langle\Psi|\Psi\rangle \big] \\
          = & \hspace{1mm} \substack{\mathrm{min} \\ \bm{J} \bm{F}} \hspace{1mm}
              \big[ \langle\Phi| e^{\hat{J}^\dag} \tilde{H} e^{\hat{J}} |\Phi\rangle / \langle\Psi|\Psi\rangle \big]
\label{eqn:jagp_energy}
\end{align}
where the second equality follows from using Eq.\ (\ref{eqn:doubles_as_jast}), defining $\tilde{H}=\hat{U}\hat{H}\hat{U}^\dag$,
and remembering that the AGP is invariant to orbital rotations.
$\bm{K}$ is found by a quasi-Newton L-BFGS \cite{Nocedal:1980:lbfgs} minimization of Eq.\ (\ref{eqn:jagp_energy}), during
which minimizations over $\bm{J}$ and $\bm{F}$ are carried out by the variational Monte Carlo (VMC)
linear method (LM) \cite{Neuscamman:2013:jagp,UmrTouFilSorHen-PRL-07,TouUmr-JCP-08}
for each value of $\bm{K}$ that is tried.
We use adjoint algorithmic differentiation \cite{Griewank-Walther-book,Neuscamman:2013:jagp} to evaluate analytic derivatives
of the resulting VMC estimates for $E(\bm{K})$ with respect to the transformed one- and two-electron integrals \cite{Helgaker_book}
that define $\tilde{H}$ and to propagate these derivatives backwards through the integral transformation and
Eq.\ (\ref{eqn:antiherm_form}) to produce the gradient $\partial E(\bm{K}) / \partial K_{ij}$.
Note that the Hellman-Feynman theorem guarantees that these $\tilde{H}$-routed derivatives, which like the energy may be evaluated
for an $O(N^5)$ cost, \cite{Neuscamman:2013:jagp} are the only nonzero contributions to $\partial E(\bm{K}) / \partial K_{ij}$.
Thus the CJAGP can be optimized variationally and at polynomial cost.

\begin{figure}[t]
\includegraphics[width=8.5cm]{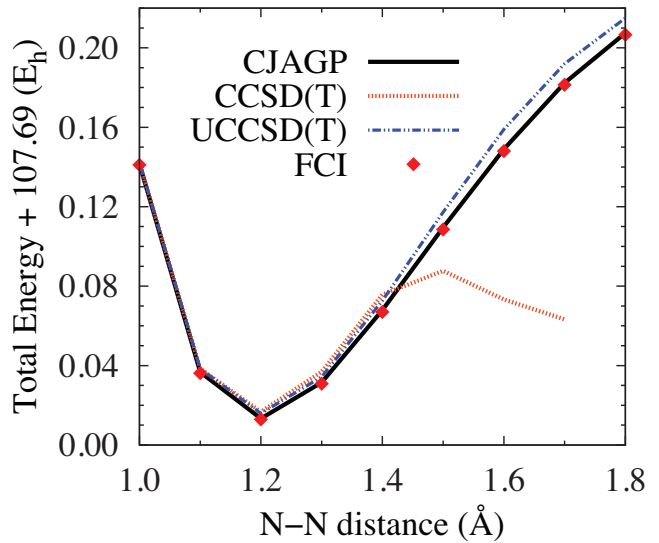}
\caption{Absolute energies in Hartrees for stretching the N$_2$ triple bond in the STO-3G basis.
         Statistical uncertainties are smaller than the line width.
        }
\label{fig:n2_sto3g_energy}
\end{figure}

\begin{figure*}[t]
\includegraphics[width=\textwidth]{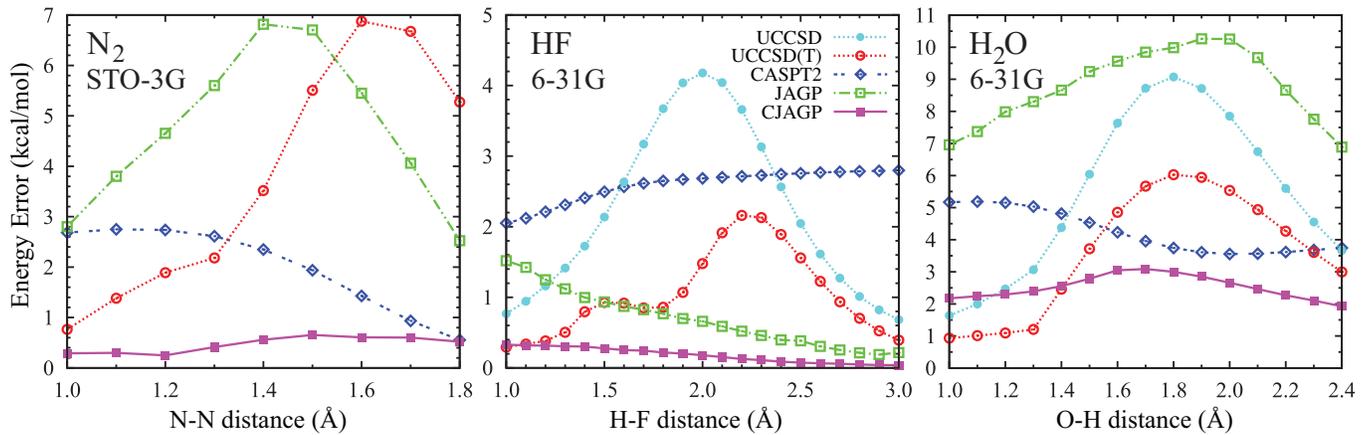}
\caption{Energy errors vs FCI for N$_2$ in STO-3G, HF in 6-31G, and symmetrically stretched H$_2$O ($\angle$HOH = 109.57$^\circ$) in 6-31G.
         Statistical uncertainties are smaller than the symbols.
        }
\label{fig:together}
\end{figure*}

Note that we use correlated sampling \cite{Umrigar:2000:corr_samp} to prevent statistical
uncertainties from upsetting the finite differences within L-BFGS.
While this approach has been sufficient for the current study, it
requires sample sizes 10 to 100 times larger than for the LM alone because
correlated sampling is less efficient when the two wave functions differ significantly, a
common occurrence during L-BFGS line searches.
In the future, we therefore hope to avoid L-BFGS and correlated sampling entirely by incorporating
the optimization of $\bm{K}$ directly into the LM.

Before discussing results, we describe the computational methods.
One- and two-electron integrals \cite{Helgaker_book} for CJAGP 
came from Psi3, \cite{Psi3}
traditional CC results from QChem, \cite{QChem:2006,QChem:2013}
and full configuration interaction (FCI) and complete active space perturbation theory (CASPT2)
results from
\uppercase{MOLPRO}. \cite{MOLPRO_brief}
CASPT2 active spaces were (6e,6o), (2e,2o), and (6e,5o) for N$_2$, HF, and H$_2$O, respectively.
Correlated methods froze non-hydrogen 1s orbitals.

\begin{table}[b]
  \caption{\% correlation recovered for 1 \AA~bonds in 6-31G.}
  \label{tab:perc_corr}
  \begin{tabular}{l r r}
    \hline
    Method  & \hspace{6mm} HF   \hspace{6mm} & \hspace{4mm} H$_2$O \\
    \hline
    CJRHF   & \hspace{6mm} 94.4 \hspace{6mm} & \hspace{4mm} 95.7 \\
    CJAGP   & \hspace{6mm} 99.6 \hspace{6mm} & \hspace{4mm} 97.5 \\
    CCD     & \hspace{6mm} 98.5 \hspace{6mm} & \hspace{4mm} 97.6 \\
    CCSD    & \hspace{6mm} 99.1 \hspace{6mm} & \hspace{4mm} 98.1 \\
    CCSD(T) & \hspace{6mm} 99.7 \hspace{6mm} & \hspace{4mm} 98.9 \\
    \hline
  \end{tabular}
\end{table}


To distinguish the effects of the CJ decomposition from those of the AGP, we have first tested a restricted Hartree-Fock (RHF) reference.
This CJRHF ansatz is produced by variationally optimizing $\bm{J}$ and $\bm{K}$ while holding $|\Phi\rangle$
fixed as the RHF determinant, which can be written as an AGP.
We see in Table \ref{tab:perc_corr} that CJRHF captures about 95\% of the correlation in
HF and H$_2$O at bond lengths near equilibrium in the 6-31G basis. \cite{POPLE:1972:6-31g_basis}
Thus, while not perfect, the CJ decomposition provides a good approximation to the full doubles
operator despite its lightweight parameterization.
This similarity suggests that we may interpret CC in these molecules through the JF language of
suppressing the simultaneous occupation of neighboring local orbitals.
This intuitive local picture of how CC encodes correlation offers an interesting complement to the traditional
analyses \cite{BARTLETT:2007:cc_review} in terms of perturbation theory and diagrammatic summation.


To test CJAGP's handling of strong correlation, we applied
it to the N$_2$ triple bond stretch in a minimal STO-3G basis. \cite{Pople:1969:sto-3g}
As seen in Table \ref{tab:npe} and Figures \ref{fig:n2_sto3g_energy} and \ref{fig:together}, CJAGP provides an excellent description
of this notoriously difficult strong correlation, \cite{BARTLETT:2007:cc_review,Knowles:2010:vcc}
with a non-parallelity error (NPE) (i.e. the difference between its highest and lowest errors relative to FCI)
significantly smaller than those of JAGP, UCCSD(T), and CASPT2.

\begin{table}[b]
  \caption{NPE (kcal/mol) for Figure \ref{fig:together}.}
  \label{tab:npe}
  \begin{tabular}{l r r r}
    \hline
    Method   & \hspace{4mm} N$_2$ &  \hspace{4mm} HF  & \hspace{4mm} H$_2$O \\
    \hline
    UCCSD(T) &  6.1  & 2.1  &  5.1   \\
    CASPT2   &  2.2  & 0.8  &  1.6   \\
     JAGP    &  4.3  & 1.3  &  3.4   \\
    CJAGP    &  0.4  & 0.3  &  1.5   \\
    \hline
  \end{tabular}
\end{table}


To test our central hypothesis that the CC, JF, and AGP components of CJAGP will work together to accurately describe
both weak and strong correlation, we applied it to bond stretching in
HF and H$_2$O in the 6-31G basis, where both strong and a substantial amount of weak correlation will be present.
Figure \ref{fig:together} and Table \ref{tab:npe} show that in both examples, CJAGP provides
a high-accuracy description of both types of correlation, outperforming UCCSD(T), JAGP, and CASPT2.
These accuracies are especially remarkable considering the compactness of CJAGP:  it contains
$O(N^2)$ variables, compared to $O(N^4)$ for UCCSD and $O(N^4+e^{N_{act}})$ for CASPT2, where $N_{act}$ is the active space size.

Inspecting the optimal transformations $\hat{U}$ reveals that the local orbitals resemble
atomic hybrids:  sp for N$_2$ and HF and sp$^3$ for H$_2$O.
We thus interpret that the CJ operator partitions molecular orbitals into chemically relavant local orbitals
an then penalizes unfavorable pairwise occupations.
This distills the correct superposition of resonance structures from the AGP, \cite{Neuscamman:2013:jagp}
which we believe is essential for CJAGP's description of strong correlation.



We have presented an operator that is simultaneously a simplified cluster operator and a Jastrow factor
compatible with quantum Monte Carlo methods and a geminal power reference function.
In contrast to traditional coupled cluster approaches, the resulting CJAGP ansatz may be optimized variationally
for a cost that scales polynomially as the fifth power of the system size, albeit with a large prefactor due to
the difficulty in combining QMC with quasi-Newton methods.
Preliminary tests show that CJAGP accurately describes both weak and strong electron correlation
in small molecules, outperforming traditional coupled cluster and complete active space perturbation theory.
Furthermore, the effectiveness of the cluster-Jastrow operator as a replacement for the
the full doubles operator reveals strong similarities between how Jastrow factors and cluster operators
handle correlation.


We thank Martin Head-Gordon for many insightful discussions and for computational resources.
We thank the Miller Institute for Basic Research in Science for funding.

%
%

\bibliographystyle{aip}
\bibliography{jastrow_cc}

\end{document}